\documentstyle[11pt,newpasp,twoside]{article}
\def\edcomment#1{\iffalse\marginpar{\raggedright\sl#1\/}\else\relax\fi}
\reversemarginpar

\begin{document}
\title{Some clues on the tilt of the Horizontal Branches in metal-rich
clusters}

\author{Raimondo$^1$ G., Castellani$^2$ V., Cassisi$^1$ S., Brocato$^1$ E., Piotto$^3$ G.}
\affil{$^1$INAF-Osservatorio Astronomico di Collurania -- Teramo,
Italy}
\affil{$^2$INAF-Osservatorio Astronomico di Monte Porzio, Roma,
Italy}
\affil{$^3$Universit\`a di Padova, Padova, Italy}

\begin{abstract}
We discuss some clues on the large tilt observed in the Horizontal
Branch of the metal-rich galactic globular clusters NGC 6388 and NGC
6441. This not yet understood feature is investigated from theoretical
and observational sides.
\end{abstract}

\section{Introduction}

Recently Brocato et al. (1999) discussed the presence of a tilt of the
order of $\Delta V \simeq 0.1\, mag$ in the HB morphology of the
intermediate metallicity globular cluster NGC 6362 [$\Delta V$ is the
magnitude difference between the top of the blue HB and the fainter
magnitude reached by the red HB (RHB)].  On the other hand, much larger
tilts ($\Delta V \simeq 0.5\, mag$) are observed in the metal rich 
clusters NGC 6388 and NGC 6441 of the inner Milky Way (Rich et
al. 1997; Sweigart and Catelan 1998).

In the present work we investigate evolutionary predictions concerning
the Color-Magnitude Diagram (CMD) of metal rich HB stars and arguments
that can constraint possible explanations of this not yet understood
feature.

\section{The theoretical side}

We computed Zero-Age Horizontal Branch (ZAHB) by adopting $Z= 0.002$ and
$Z= 0.006$ as representative of "metal rich" clusters like 47 Tuc, and
$Z=0.02$ as a safe upper limit for globular cluster metallicities. As
expected, the effective temperature of RHB stars depends on the mixing
length assumption: at any given temperature, the luminosity of the
red-ZAHB increases when the mixing length increases. At higher
metallicities the HB locus moves toward lower effective temperatures
and RHB stars are increasingly affected by external convection.

Canonical ZAHBs with metallicity larger than $[M/H]=-1$ should be
affected by a tilt, that increases by: $1)$ increasing the
metallicity; and $2)$ decreasing the mixing length value. The larger
reasonable slope is: $dM_V / d(B-V) \simeq 0.2$ for $ Z_{\sun}$ and
$\alpha =1.6$, unfortunately not large enough to account for the
observed tilt.
Un-canonical frameworks do not seem to enlighten the
problem (Raimondo et al. 2002).

\section{The observational side}

The CMDs of NGC 6441 and NGC 6388 (Fig. 1) show that the lower
envelope of the HB is sloped as $dV / d(B-V) \sim 1.5$. We tried to
explain this feature as a result of {\it differential reddening } or
as due to a peculiar {\it metallicity spread}. Our attempts were not
conclusively successful (Raimondo et al. 2002).  Moreover, if one
analyzes the CMDs of the metal rich globular clusters in the Padova HST
snapshot data-base (Piotto et al. 2002) it can be seen (Fig. 1) that
\emph{ the clusters more metal rich than 47 Tuc show a tilted RHB}
with a slope $dV/d(B-V) \sim 1.5$.

\section{Conclusions}

The large tilt of the RHB of NGC 6441 and NGC 6388 seems to be a
feature observed in many metal rich clusters. Up to now, a fully
satisfactory explanation is not yet found. More precise observations
(i.e. FLAMES@VLT) are urgently needed to properly constraint our
understanding of metal rich stellar systems.

\begin{figure}[t]
\plotfiddle{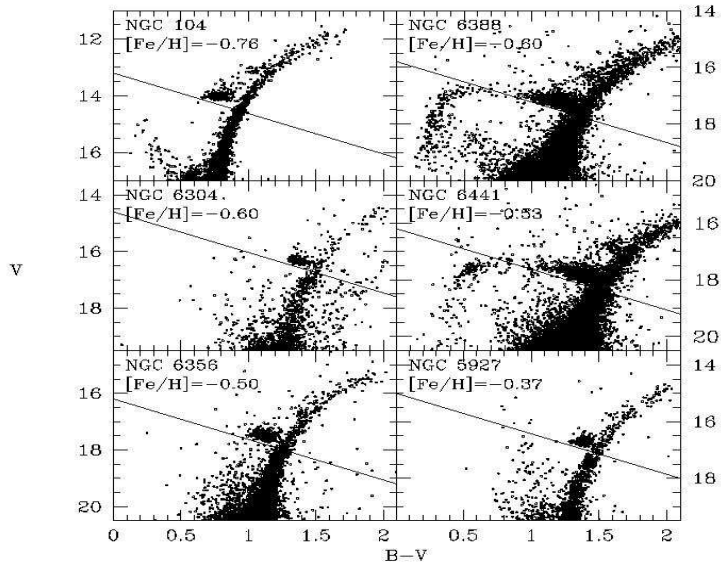}{0in}{0}{50}{40}{-150}{-280} \vspace{7.5cm}
\caption{HBs of six well populated clusters in Piotto et al.  (2002)
data-base.  The solid line is the slope $dV / d(B-V)= 1.5$ (see
text). }
\end{figure}


\begin{references}

\reference{Brocato, E., Castellani, V., Raimondo, G.  \& Walker, A. R., 1999, ApJ, 527, 230}
\reference{Piotto G., et al. 2002,  A\&A, 391, 945}
\reference{Raimondo G., Castellani V., Cassisi S., et al. 2002, ApJ, 569, 975}
 \reference{Rich et al. 1997, ApJ,
484, L25}
\reference{Sweigart, A. V. \& Catelan, M., 1998, ApJ,
501, L63}
\end{references}
\end{document}